\documentclass[useAMS,usenatbib]{mn2e}
\usepackage{graphicx}
\usepackage{epsfig}

\title[The first optical study of G182.4+4.3]{The first optical light from the supernova remnant G182.4+4.3 located in the
Galactic anti-centre region}
\author[A. Sezer, F.~G\"{o}k, E.~Aktekin]{A. Sezer,$^{1,2}$\thanks{E-mail: aytap.sezer@boun.edu.tr (AS); gok@akdeniz.edu.tr
(FG); eaktekin@akdeniz.edu.tr (EA)} F.~G\"{o}k$^{3}$ and
E.~Aktekin$^{4}$ \\
$^{1}$T\"UB\.ITAK Space Technologies Research Institute, ODTU
Campus, Ankara, 06531, Turkey\\
$^{2}$Bo\~gazi\d{c}i University, Faculty of Art and Sciences,
Department of Physics, \.Istanbul, 34342, Turkey\\
$^{3}$Akdeniz University, Faculty of Education, Antalya, 07058, Turkey\\
$^{4}$Akdeniz University, Faculty of Sciences, Department
of Physics, Antalya, 07058, Turkey\\
}
\begin{document}

\date{}

\pagerange{\pageref{firstpage}--\pageref{lastpage}} \pubyear{2012}

\maketitle

\label{firstpage}

\begin{abstract}
We report the discovery of optical filamentary and diffuse
emission from G182.4+4.3 using 1.5-m Russian-Turkish telescope. We
present the optical CCD images obtained with H$\alpha$ filter
revealing the presence of mainly filamentary structure at the
northwest, filamentary and diffuse structure at the centre, south
and north regions of the remnant. The bright optical filaments
located in northwest and south regions are well correlated with
the prominent radio shell of the remnant strongly suggesting their
association. From the flux-calibrated CCD imaging, the average
[S\,{\sc ii}]/H$\alpha$ ratio is found to be $\sim$0.9 and
$\sim$1.1 for south and northwest regions, which clearly indicates
that the emission originates from the shock heated gas. We also
present the results of X-ray data obtained from {\it XMM-Newton}
that show diffuse emission with a very low luminosity of
$\sim$$7.3\times10^{31}$ erg s$^{-1}$ at a distance of 3 kpc in
0.3$-$10 keV energy band. Furthermore, we find a surprisingly
young age of $\sim$4400 yr for this remnant with such a large
radius of $\sim$22 pc.

\end{abstract}


\section{Introduction}

There are 274 Galactic supernova remnants (SNRs) catalogued in
\citet {b12} by their radio emission. Most of them are located in
the Galactic plane where the density of gas and dust is very high
which make the observation, except for radio band, of SNRs hard
due to strong interstellar extinction and reddening effects in
their line of sight. This difficulty can be surmounted by using
narrow-band filters such as H$\alpha$, [S\,{\sc ii}], [N\,{\sc
ii}] and [O\,{\sc iii}] centered on characteristic emission lines
on deep exposures. A number of Galactic SNRs are detected
especially with H$\alpha$ filter (e.g. \citet {b9, b6, b8, b3,
b4}), in which they usually have peculiar morphological
structures. Some of them have filamentary structure while others
have filamentary and diffuse structures together or arc (shell)
structures. \citet {b5} reported a catalogue of 24 known Galactic
SNRs uncovered in H$\alpha$ light in the Anglo-Australian
Observatory/United Kingdom Schmidt Telescope (AAO/UKST) H$\alpha$
survey of the southern Galactic plane. The optical observations of
Galactic SNRs enable us to study the physical conditions in the
remnant and the ambient medium, such as variation of chemical
composition, density and evolutionary state.

Galactic SNR G182.4+4.3 located in the anti-centre region has been
detected at 1400 MHz, 2675 MHz, 4850 MHz, and 10450 MHz with the
Effelsberg 100-m telescope in radio bands by \citet {b1}. They
reported that the remnant has a shell structure with a radio
spectral index of $\alpha$=$-0.42\pm0.10$
(S$_{\nu}\sim\nu^{\alpha}$), about 50 arcmin in size, very low
radio surface brightness (7.5$\times 10^{-23}$ Watt m$^{-2}$
Hz$^{-1}$ sr$^{-1}$ at 1 GHz), and is in Sedov expansion stage.
Expanding into an ambient medium of low density indicated that its
location was in front of the outer spiral arm III. H\,{\sc i} line
observations of neutral hydrogen showed that the column density
towards the remnant must be $\leq$4$\times$10$^{21}$ cm$^{-2}$.
Considering its low ambient density and distance to be 3 kpc, they
argued that z-height of the remnant was about 230 pc and concluded
that it was most likely the remnant of a Type Ia supernova
explosion. They have searched the {\it ROSAT} all sky survey at
the position of this remnant and detected no visible X-ray
emission (corresponding upper limit of 6.2$\times 10^{-2}$ counts
s$^{-1}$). Assuming an initial explosion energy of $10^{51}$ erg,
they obtained ambient density $n_{0}\sim 0.013$ ${\rm cm}^{-3}$,
swept-up mass $M_{\rm sw}\sim 14$ M\sun, electron temperature
$kT\sim 5.0$ keV, age 3800 yr, and shock velocity $V_{\rm
exp}=2300$ km ${\rm s}^{-1}$.

There is no optical identification of G182.4+4.3 in literature so
far. In this paper, we report the first detection of optical
emission from G182.4+4.3 both filamentary and diffuse structure in
H$\alpha$. We also studied {\it XMM-Newton} data to investigate
its properties in X-ray bands and combine optical imaging and
X-ray observation to present a better view of this SNR.

The paper is organized as follows: Observations and data reduction
are described in Section 2. Based on the CCD imaging analysis
together with X-ray results, we discuss the filamentary and
diffuse structure and the plasma parameters of the SNR in Section
3. We give our conclusion in Section 4.

\section[]{Observations and data reduction}
\subsection{Optical Imaging}
Optical CCD imaging observations of G182.4+4.3 were taken on 2011
February 05, March 02 and 2012 February 19, with low resolution
spectrograph TFOSC (TUG Faint Object Spectrograph and Camera)
equipped with  2048$\times$2048 back-illuminated camera with a
pixel size of 15 $\mu m$ $\times$ 15 $\mu m $ in a
13.5$\times$13.5 arcmin$^{2}$ field of view (FOV) attached to the
Cassegrain focus of the 1.5-m Russian-Turkish joint telescope
(RTT150)\footnote{Details on the telescope and the spectrograph
can be found at http://www.tug.tubitak.gov.tr.} at T\"{U}B\.{I}TAK
National Observatory (TUG) T\"{u}rkiye, Antalya. The optical
images were obtained with H$\alpha$, [S\,{\sc ii}] and their
continuum filters. The characteristics of these
interference-filters are summarized in Table 1. The images were
reduced by using standard {\sc iraf} (Image Reduction Analysis
Facility) routines for background subtraction, flat-fielding,
trimming and continuum subtraction. The spectroscopic standard
star HR5501 \citep {b10, b11} was used for the absolute flux
calibration. We write the coordinate information into the FITS
header of the individual images with world coordinate system (WCS)
tools. The weather conditions during these imaging observations
were poor.

Due to its large size, we divided the whole remnant into several
fragments and observed each fragment with H$\alpha$ filter with
short exposure time of 100 s. We obtained optical emission only
from four regions, namely north (N), northwest (NW), center (C)
and south (S). We focused our study on these four regions to
obtain images with H$\alpha$, [S\,{\sc ii}] and their continuum
filters with relatively long exposures (900 s). Continuum
subtracted H$\alpha$ images of NW, S, C and N regions are given in
Fig. 1, 2 and 3.

We also took the long-slit spectra by locating the slit on the
brightest filament in S region on 2011 November 07 with an
exposure time of 7200 s, with the spectroscopic mode of the TFOSC
attached to Cassegrain focus of the RTT150. Unfortunately, due to
poor observing conditions the obtained spectra did not allow us to
get reliable results.

\subsection{{\bf X-ray and radio-continuum observations}}
{\it XMM-Newton} \citep {b32} observed G182.4+4.3 on 2001 Jun 15,
under the observation ID 50406540 and exposure time of 23 ks. {\it
XMM-Newton} has three X-ray telescopes \citep {b33}, one equipped
with EPIC-PN \citep {b30} and two with EPIC-MOS \citep {b31} CCD
detectors in the focal plane. The data were reduced and analyzed
using the {\sc xmm-newton sas}\footnote{Science Analysis Software
({\sc sas}), http://xmm.vilspa.esa.es/sas/}, version 1.52.8.
Calibrated event files for the EPIC-MOS1, EPIC-MOS2 and EPIC-PN
detectors were produced using {\sc sas} task {\sc emchain} and
{\sc epchain} and following standard procedures. EPIC-MOS2 image
in 0.3$-$10 keV energy band of the G182.4+4.3 is given in Fig. 4.

The overall radio structure of the G182.4+4.3 at 4850 MHz with the
Effelsberg 100-m telescope (R. Kothes, private communication) is
given in Fig. 5. As seen from the figure, the radio shell is most
prominent in the south-west direction, this is because the
prominent part of the shell is expanding towards the Galactic
plane into higher density medium while the top shell is expanding
away from the plane into very low density medium \citep {b2}.

We overlaid the H$\alpha$ mosaic image of the observed regions
with radio-continuum contour image taken from 4850 MHz Effelsberg data 
and X-ray contour image from {\it XMM-Newton} data to see if there is an association between
the optical, radio and X-ray emission. As seen from the Fig. 6,
there is a good correlation between optical and radio emission in
NW and S regions. Due to poor observing conditions, for N and C
regions the correlation is not seen clearly. The optical emission
of S region correlates with both the radio and X-ray emission. N,
NW and C regions are stand out of the FOV of {\it XMM-Newton}.

\section{Results and Discussion}

We present the first CCD images of G182.4+4.3 together with the
results from {\it XMM-Newton} archival data analysis.

The H$\alpha$ image of NW region shows mainly filamentary
structure while that of other regions show both filamentary and
diffuse structure, see Fig. 1, 2 and 3. As seen from Fig. 1,
strong multiple filaments overlapping with the surrounding diffuse
emission are visible only in the lower left part of NW region. In
S region, see Fig. 2, a strong and long filamentary structure
extending from the southeast edge of the field up to the northwest
edge is visible. There is also a weaker filament extending
parallel to the previous one in the southwest edge of this region.
In the upper part of the prominent filament a strong diffuse emission is
visible. In C region (see Fig. 3, upper panel), there is a curved
weak filament extending from south to northwest as well as a
fainter diffuse emission. The reason of the weakness of the
emission may be the presence of very bright stars in this field. N
region (see Fig. 3, bottom panel) seems much complex compared to
other regions in H$\alpha$ emission. There are several small scale
filaments embedded in diffuse emission in this region. The
filaments seen in the NW and S regions are very well correlated
with the prominent radio shell of the remnant.

The [S\,{\sc ii}]/H$\alpha$ ratio greater than 0.5 is used as a
standard discriminator for shock heated gas and hence indicative
of SNRs \citep {b40}. We obtained [S\,{\sc ii}]/H$\alpha$ ratio to
be 0.9$\pm$0.1 and 1.1$\pm$0.1 for S and NW regions, respectively.
These ratios indicate that the emission originates from the
ionization of shock heated gas resulted from collision.

We detected a diffuse and faint X-ray emission coming from the
remnant as seen from the Fig. 4 and obtained a
background-subtracted count rate of 0.04 counts s$^{-1}$. Even
with this low-count X-ray detection, we tried to get some
information on the X-ray nature of G182.4+4.3 using the EPIC-MOS1
spectrum. We extracted the spectrum from a circular region with a
radius of 8.3 arcmin centered at $\rmn{RA}(2000)=06^{\rmn{h}}
07^{\rmn{m}} 10^{\rmn{s}}$, $\rmn{Dec.}~(2000)=28\degr 52\arcmin
05\arcsec$. The spectrum was grouped with a minimum of 30 counts
bin$^{-1}$ and we used $\chi^{2}$ statistics. The spectrum was
best-fitted with VPSHOCK model which is suitable for modelling
plane-parallel shocks in young SNRs where plasma has not reached
the ionization equilibrium \citep {b27}, modified by Galactic
absorption via the WABS multiplicative model \citep {b13} using
{\sc xspec} v11.3 \citep {b34}. While model fitting, the absorbing
column density ($N_{\rm H}$), electron temperature ($kT_{\rm e}$)
and ionization parameter ($\tau=n_{\rm e}t$) were set free while
all elements were fixed at their solar abundances \citep {b26}.
This fitting gave us $N_{\rm H}$ $\sim 5\times 10^{21}$ cm$^{-2}$,
$kT_{\rm e}$ $\sim 0.9$ keV, $\tau$ $\sim 0.3\times 10^{10}$
cm$^{-3}$s and a low X-ray luminosity ($L_{\rm x}$) in 0.3$-$10
keV energy band to be $\sim$$7.3\times 10^{31}$ erg s$^{-1}$ at a
distance of 3 kpc with an acceptable reduced $\chi^{2}$ value of
1.11 (64.5/58 d.o.f.). Using the emission measure ($EM=n_{\rm
e}n_{\rm H}V$, where $n_{\rm e}$ and $n_{\rm H}$ are number
densities of electrons and protons respectively, and $V$ is the
X-ray emitting volume, and assuming $n_{\rm e}=1.2n_{\rm H}$) we
obtain a significantly low electron density $n_{\rm e}$ of
$\sim$0.024 $\rm cm^{-3}$. The mass of the X-ray emitting gas is
calculated to be $\sim$$1{\rm M}\sun$ from $M_{\rm x}=m_{\rm
H}n_{\rm e}V$, where $m_{\rm H}$ is mass of a hydrogen atom. From
t=$\tau$/$n_{\rm e}$ the age of G182.4+4.3 is calculated to be
$\sim$4400 yr which is consistent with the age that \citet {b1}
found from {\it ROSAT} data. From the size distribution of the
SNRs in the Magellanic Clouds, \citet {b42} showed that the
distributions of close to uniform between r$\sim$10 pc and
r$\sim$30 pc. They argued that a uniform size distribution arises
from the physics of SNR evolution, Sedov expansion model for SNRs,
and the distributions of densities in the ambient medium. Thus,
being at t$\sim$4400 yr age and having r$\sim$22 pc, G182.4+4.3 is
likely to be in the Sedov phase. A 4400 yr old SNR with a given
original supernova mass, can expand to such a large size in a low
density medium. The location of G182.4+4.3 is high from the
Galactic plane ($\sim$230 pc), it has a very low radio surface
brightness and is faint in X-rays indicating that the remnant is
expanding in a low density medium. The low electron density
obtained from X-ray spectral analysis which is consistent with
that of {\it ROSAT} data also supports this idea. Regarding its
large radius and young age, G182.4+4.3 resembles the shell type
Galactic SNRs G93.3+6.9 (DA530) and G156.2+5.7. G93.3+6.9 is
located at 420 pc above the Galactic plane at a distance of 3.8
kpc and has a diameter of D=26 pc \citep {b41}. It is at the age
of 5000 yr and expanding in a very low-density medium (0.05
cm$^{-3}$). It has a very low radio surface brightness and its
X-ray emission is extremely faint \citep {b41}. The other example
G156.2+5.7 is located at z$>$130 pc, at a distance of $\sim$3 kpc
and is expanding in a low-density medium (0.01 cm$^{-3}$) \citep
{b44}. Its optical observations suggest that it is interacting
with clumpy interstellar medium \citep {b43}. It has low radio
surface brightness however, it is bright in X-rays. \citet {b1}
compared G182.4+4.3 with the G156.2+5.7 in this respect. They
argued that the difference in the X-ray luminosities of G182.4+4.3
and G156.2+5.7 may be resulted from the difference in the ratio
between the mass of the clouds and the mass of the intercloud
medium. They concluded that G182.4+4.3 expanded into a medium with
low density and a low fraction of clouds. Propagation of an SNR
shock wave into a cool dense interstellar medium leads to
H$\alpha$ emission. The filamentary H$\alpha$ emission emanating
from this remnant suggests that there might be clumps of clouds in
the medium in which SNR is expanding. Furthermore the different
scale and structures of the filaments together with the diffuse
emission may indicate the presence of small or large scale
inhomogeneities in the interstellar clouds. However, in the FOV of
{\it XMM-Newton} (see Fig 4), the cloud density might be very low
leading to a faint X-ray emission.

From the optical spectra of SNRs, the electron density can be
estimated by using the flux ratio of [S\,{\sc
ii}]($\lambda$6716/$\lambda$6731) \citep {b21}. The estimated
electron density leads to obtain the pre-shock cloud density (see
\citet {b7}). Future spectral observations of this remnant could
give detailed information about the inhomogeneities in the SNR and
its ambient medium.

\section{Conclusion}

We detected optical filamentary and diffuse emission for the first
time for G182.4+4.3 with several short or long scale filaments found
in the N, NW, C and S regions. The strong optical
filaments located in NW and S regions match with the prominent
radio shell at 4850 MHz Effelsberg data of the remnant suggesting
their association. A very strong emission of [S\,{\sc ii}]
relative to H$\alpha$ [[S\,{\sc ii}]/H$\alpha \sim$0.9, $\sim$1.1]
obtained from imaging suggests that the emission originates from
the shock heated gas. Finally, we estimated X-ray properties of
this remnant using {\it XMM-Newton} archival data. The X-ray spectrum of
G182.4+4.3 shows that the plasma is of thermal origin in
non-equilibrium ionization state and requires a temperature of
$\sim$0.9 keV. The best-fitting of the spectrum implies that the
SNR is very young ($\sim$4400 yr), expanding in a very low density
medium ($\sim$0.024 $\rm cm^{-3}$), it has a low X-ray emitting
mass ($\sim$$1{\rm M}\sun$) and very low X-ray luminosity
($\sim$$7.3\times 10^{31}$ erg s$^{-1}$ in 0.3$-$10 keV).

\section*{Acknowledgments}
We are grateful to M. Filipovic for his constructive comments and
suggestions on the manuscript. We thank to T\"{U}B\.{I}TAK for a
partial support in using RTT150 (Russian-Turkish 1.5-m telescope
in Antalya) with project number 09ARTT150-458-0. We also thank to
A. Aky\"{u}z for providing us some of the interference filters and
R. Kothes for providing us the 4850 MHz Effelsberg data. The authors
acknowledge support by the Akdeniz University Scientific Research
Project Management. AS is supported by T\"{U}B\.{I}TAK
PostDoctoral Fellowship.

\onecolumn

\begin{table*}
 \begin{minipage}{140mm}
  \caption{The characteristics of interference
  filters and exposure times of imaging observations taken with TFOSC CCD.}
 \begin{tabular}{@{}ccc@{}}
  \hline
     Filter & Wavelength & Exposure times (s) \\
& (FWHM) ($\AA$)&  \\
 \hline
H$\alpha$& 6563 (88) &  900\\
H$\alpha$ cont.& 6446 (130) & 900\\

[S\,{\sc ii}]& 6728 (70) &  900 \\

[S\,{\sc ii}] cont.& 6964 (300)& 900 \\
\hline
\end{tabular}
\end{minipage}
\end{table*}

\begin{figure}
\centering
\vspace*{1pt}
\includegraphics[width=21cm]{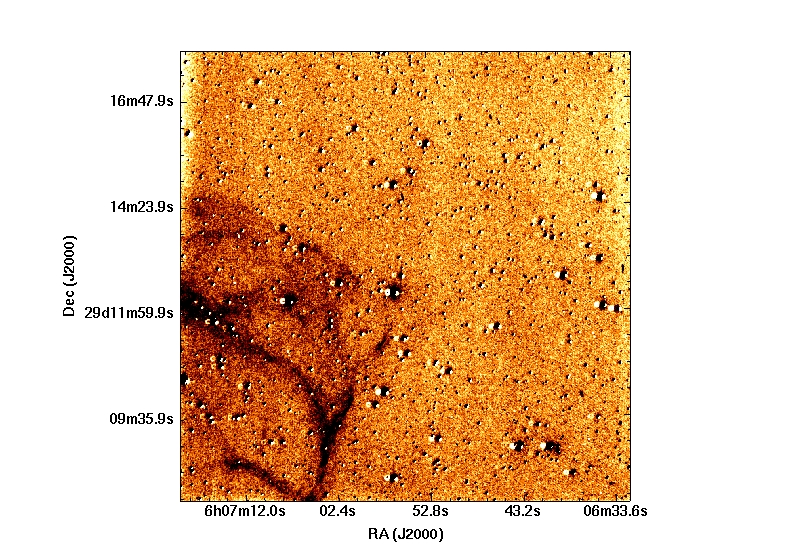}
\caption{Continuum subtracted and smoothed H$\alpha$ image of NW
region. A network structure of filaments is seen in the lower left
part of this image.}
\end{figure}

\begin{figure}
\centering
  \vspace*{1pt}
\includegraphics[width=15cm]{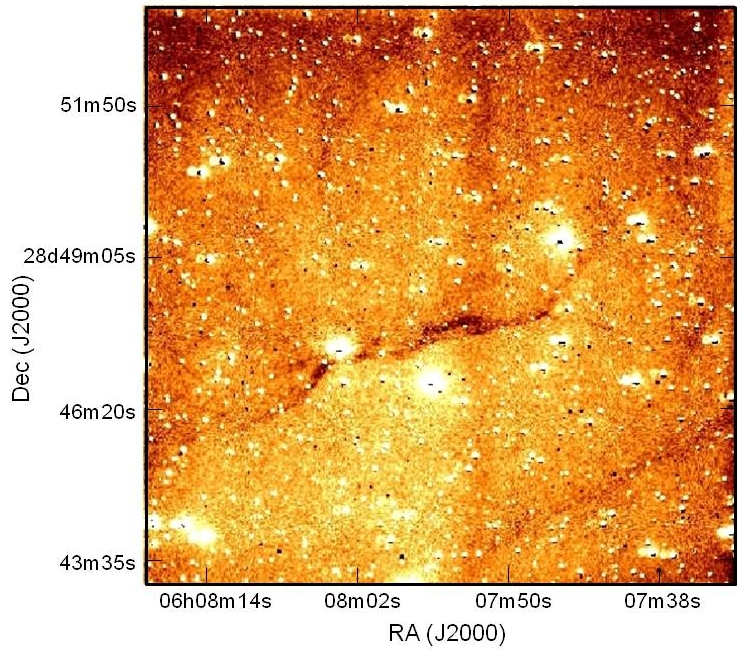}
\caption{Continuum subtracted and smoothed H$\alpha$ image of S
region. A long and sharp filament, which is $\sim$8 arcmin extend,
is noticeable in this figure.}
\end{figure}

\begin{figure}
\centering
\includegraphics[width=12.5cm]{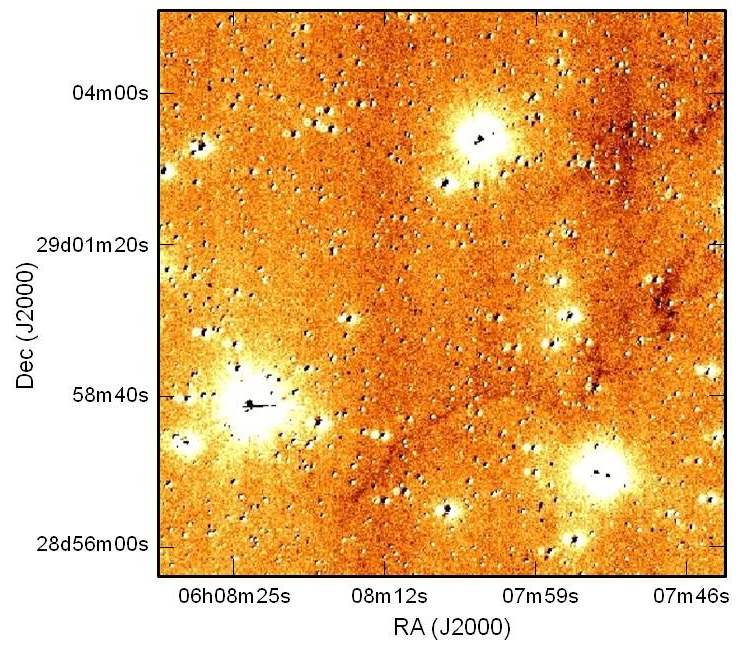}
\includegraphics[width=15.5cm]{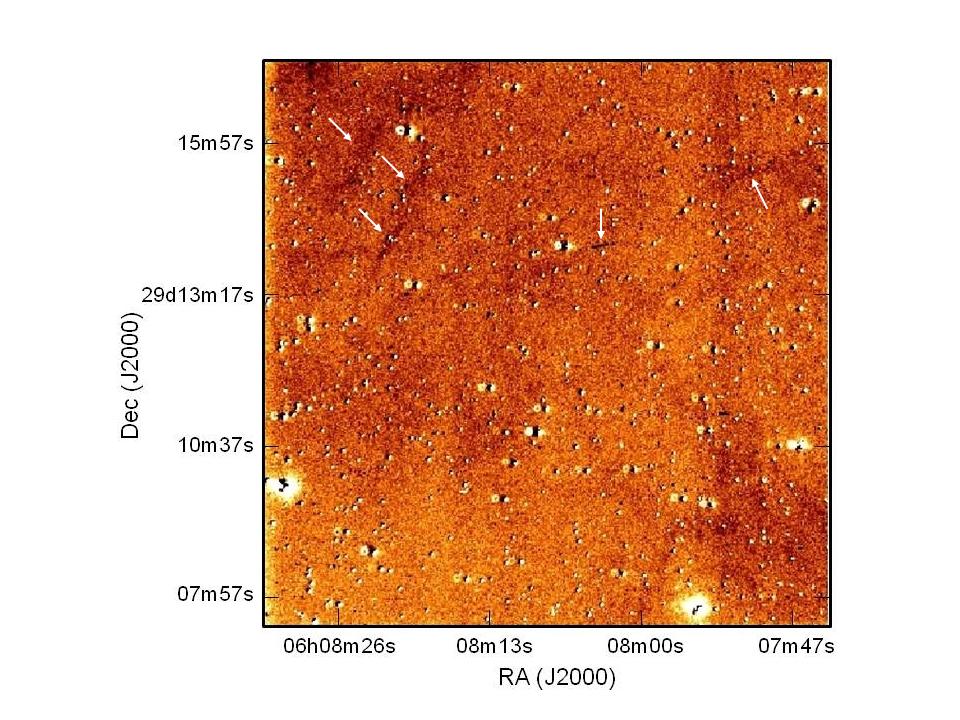}
\caption{The upper and bottom panels show continuum subtracted and
smoothed H$\alpha$ images of C and N regions, respectively.
Filamentary structure is noticeable in C region, while both
filamentary (pointed with arrows) and diffuse structure are
present in N region.}
\end{figure}

\begin{figure}
\centering \vspace*{17pt}
\includegraphics[width=8.5cm]{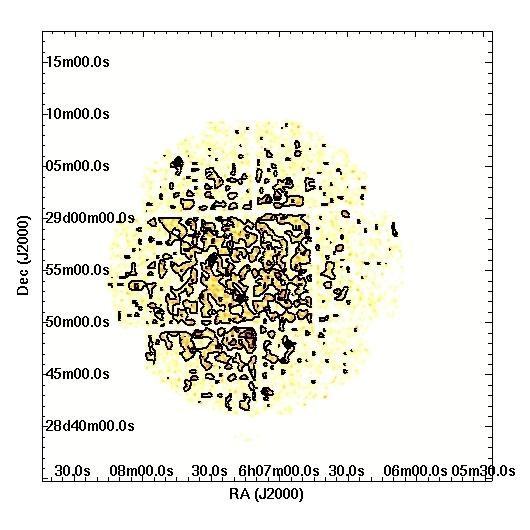}
\caption{The EPIC-MOS2 image of G182.4+4.3 in the 0.3$-$10 keV
energy band with X-ray contour levels of 2, 2.86, 4.09, 5.85 and 7
Jy beam$^{-1}$.}
\end{figure}

\begin{figure}
\centering
  \vspace*{17pt}
 \includegraphics[width=9cm]{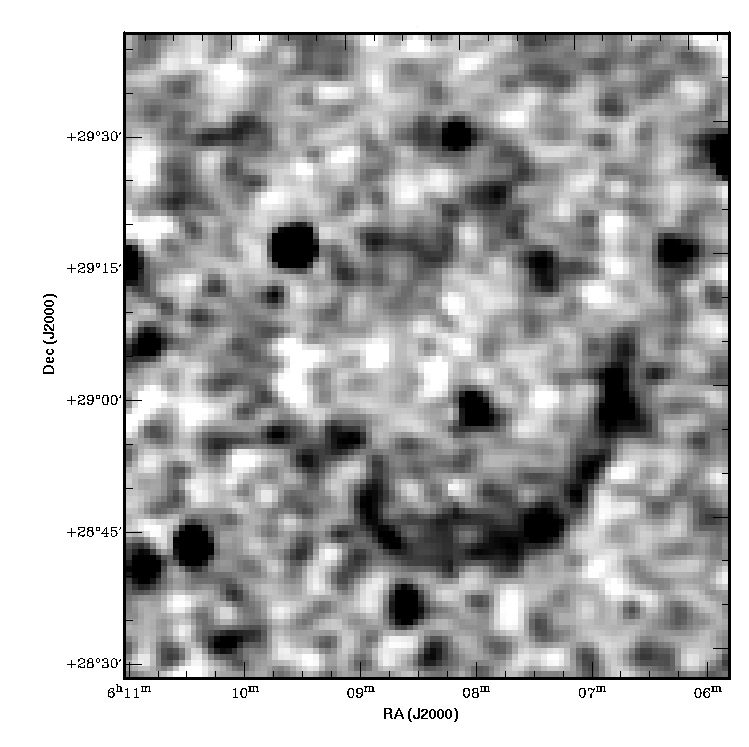}
  \caption{The radio-continuum image taken from 4850 MHz
Effelsberg data. A typical shell with a diameter of about 50
arcmin.}
\end{figure}

\begin{figure}
\centering
  \vspace*{17pt}
 \includegraphics[width=13.5cm]{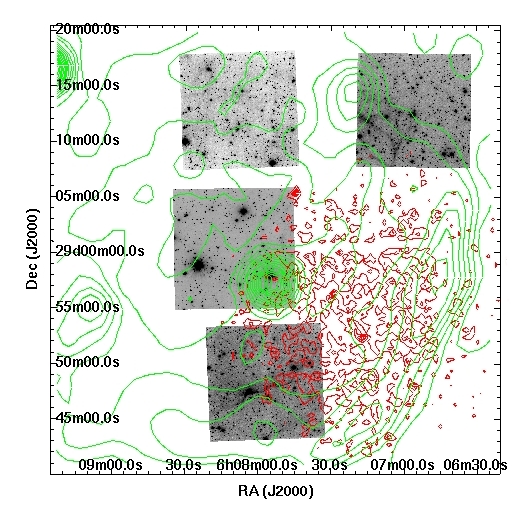}
 \includegraphics[width=7cm]{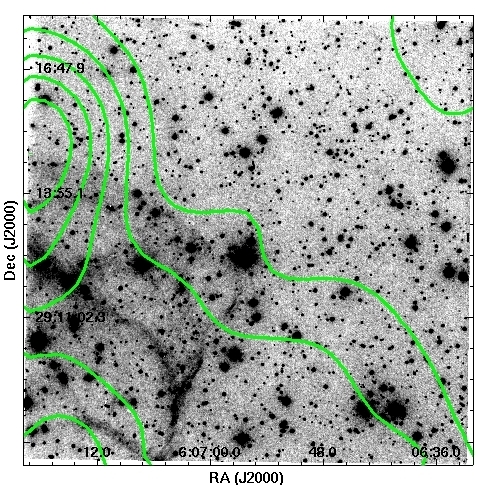}
 \includegraphics[width=7cm]{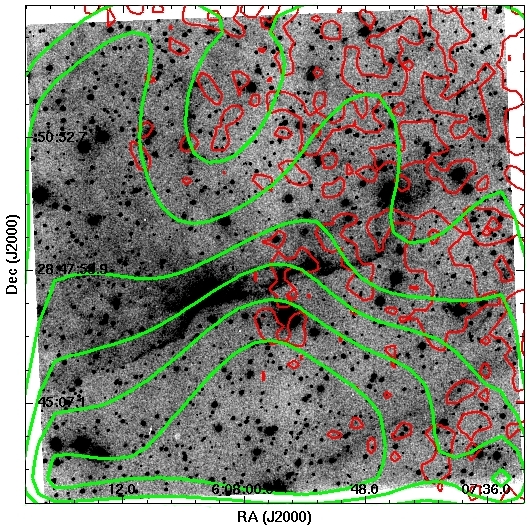}
  \caption{Upper panel: The mosaic image combines optical images (H$\alpha$) of four
  fields (N, NW, C and S) of G182.4+4.3 overlaid with (i) radio-continuum contours (green) taken from 4850 MHz
Effelsberg data and (ii) X-ray contours (red) from EPIC-MOS2. The
radio contour levels are -2.95, 1.32, 5.59, 7.73, 14.13 mJy
beam$^{-1}$ and the X-ray contour levels are 2.02, 2.86, 4.09,
5.85 and 7 Jy beam$^{-1}$. Each optical image (N, NW, C and S)
covers an area of $\sim$10$\times$10 arcmin$^{2}$. The whole
figure covers an area of 40$\times$40 arcmin$^{2}$. Lower left
panel: The enlarged H$\alpha$ image of NW region overlaid with
radio-continuum contours (green). The contour levels are -0.67,
2.94, 6.54 and 10.14 mJy beam$^{-1}$. Lower right panel: The
enlarged H$\alpha$ image of S region overlaid with radio-continuum
(green) and X-ray contours (red). The radio contour levels are
0.11, 2.09, 4.09 and 7.08 mJy beam$^{-1}$ and the X-ray contour
levels are 2.02, 2.88, 4.11 and 7 Jy beam$^{-1}$.}
\end{figure}

\twocolumn
\end{document}